\documentclass[aps,prb,twocolumn,showpacs]{revtex4}
\usepackage{graphicx}

\begin{document}

\title{Pressure induced superconductivity in Bi single crystals}

\author{Yufeng Li, Enyu Wang, Xiyu Zhu$^{*}$, and Hai-Hu Wen}
\email[Electronic address:]{ hhwen@nju.edu.cn, zhuxiyu@nju.edu.cn}
\affiliation{Center for Superconducting Physics and
Materials, National Laboratory of Solid State Microstructures and
Department of Physics, National Collaborative Innovation Center of Advanced Microstructures, Nanjing University, Nanjing 210093, China}

\begin{abstract}
Measurements on resistivity and magnetic susceptibility have been carried out for Bi single crystals under pressures up to 10.5GPa. The temperature dependent resistivity shows a semimetallic behavior at ambient and low pressures (below about 1.6GPa). This is followed by an upturn of resistivity in low temperature region when the pressure is increased, which is explained as a semiconductor behavior. This feature gradually gets enhanced up to a pressure of about 2.52GPa. Then a non-monotonic temperature dependent resistivity appears upon further increasing pressure, which is accompanied by a strong suppression to the low temperature resistivity upturn. Simultaneously, a superconducting transition occurs at about 3.92K under a pressure of about 2.63GPa. With further increasing pressure, a second superconducting transition emerges at about 7K under about 2.8GPa. For these two superconducting states, the superconductivity induced magnetic screening volumes are quite large. As the pressure further increased to 8.1GPa, we observe the third superconducting transition at about 8.2K. The resistivity measurements under magnetic field allow us to determine the upper critical fields $\mu_0 H_{c2}$ of the superconducting phases. The upper critical field for the phase with $T_c=3.92$K is extremely low. Based on the Werthamer-Helfand-Hohenberg (WHH) theory, the estimated value of $\mu_0 H_{c2}$ for this phase is about 0.103T. While the upper critical field for the phase with $T_c$=7K is very high with a value of about 4.56T. Finally, we present a pressure dependent phase diagram of Bi single crystals. Our results reveal the interesting and rich physics in bismuth single crystals under high pressure.
\end{abstract}


\maketitle

\section{Introduction}
Bismuth (Bi) is a very interesting element in terms of electronic properties. The hybridized p-orbitals of Bi exhibit tremendous and rich physics, and thus have received much attention. Firstly, the band structure calculation on Bi shows a feature of multiband and multi-Fermi pockets\cite {Biband, Bipocket, Bielc-hole}. Due to the existence of many small electron and hole Fermi pockets, it shows the giant magnetoresistance effect\cite{BiMR}. In addition, as a very heavy element, Bi is supposed to have strong spin-orbital coupling effect\cite{BiSOC}. Concerning superconductivity, early investigations gave segmental messages\cite{BiHc2,BiAmo,BiNano,Bisemi,BipressureSC}. At room temperature, Bi shows sequential structure transitions by varying the applied pressure\cite{BipressureSC}: Bi-I (ambient phase) to Bi-II at about 2.53GPa; Bi-II to Bi-III at about 2.70GPa; Bi-III to Bi-IV at about 4.48GPa; Bi-IV to Bi-V at about 6.50GPa and Bi-V to Bi-VI at about 7.98GPa. Except for the Bi-I phase, other phases mentioned above are all reported to show superconductivity at low temperatures. In addition, amorphous Bi is also superconductive in form of thin film with $T_c$ of about 6K\cite{BiAmo}. Furthermore, granular Bi nanowires fabricated by deposition show two superconducting transitions at 7.2K and 8.3K\cite{BiNano}. Besides, in the study of some single crystals grown by Bi flux, interfacial superconductivity was also observed\cite{SrMnBi}. Although these pieces of knowledge have been already accumulated in the past on Bi, the detailed pressure dependent evolution of superconductivity and magnetic properties are hard to be found in literatures. Concerning the mysterious behavior of superconducting Bi under pressure, it is necessary to measure the basic superconducting properties of Bi. In addition, nowadays high pressure has become a very powerful tool for exploring new superconductors\cite{CrAs,MnP,ZrTe5}. For the materials containing Bi element, cautions must be taken when claiming superconductivity under pressure since Bi impurity might exist in the sample. Our study will also serve as a useful reference for this purpose.

In this work, we report the temperature dependent resistivity and magnetic susceptibility under pressures up to 10.5GPa. The evolution of Bi-I to superconducting Bi-II and Bi-III phases induced by high pressure is shown by resistivity measurements. The measurements of magnetic susceptibility under high pressures reveal that the Bi-VI phase becomes superconducting below 8K at and above 8.1GPa. Magnetization hysteresis loops and superconducting phase diagram of bismuth under high pressures are also obtained.

\section{EXPERIMENTAL METHOD}
The Bi single crystals were taken from the commercial product of Bi granules (Alfa Aesar, 99.997\% purity). The samples look shinny with the typical size of about 2mm. Temperature dependent resistivity measurements were carried out with a physical property measurement system (PPMS-16T, Quantum Design). The resistive measurements at high pressures were accomplished by using the HPC-33 piston type pressure cell. DC magnetization measurements were performed with a SQUID-VSM-7T (Quantum Design). The DC magnetic susceptibility measurements at high pressures were accomplished by using the Diamond Anvil Cell (DAC) designed by the Honest Machinery Designer's office (HMD). A small piece of ruby was used as the pressure manometer. Daphne oil 7373 was used as the pressure transmitting medium in both resistive and magnetic susceptibility measurements.

\begin{figure}
\includegraphics[width=9cm]{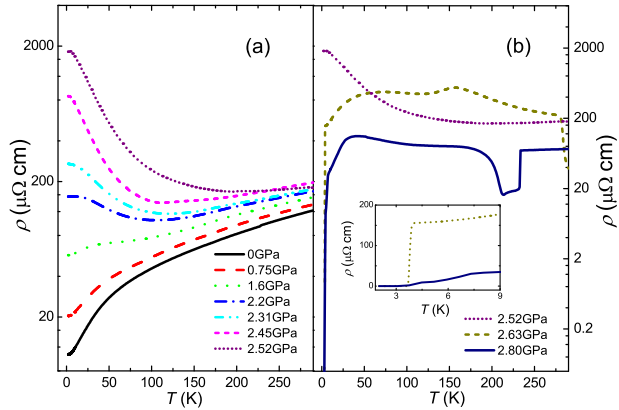}
\caption{(Color online) Temperature dependent resistivity under pressures up to 2.80GPa. Bi shows semimetal to semiconductor (see text) transition at a pressure of about 1.6GPa. Then it becomes superconducting at pressures above 2.63GPa. There are two resistivity anomalies occurring in the temperature region from 200K to 250K under 2.80GPa, which may be related to phase transitions on cooling. The inset shows the one step superconducting transition under 2.63GPa and the two step transitions under 2.8GPa.}
\label{fig1}
\end{figure}
\section{RESULTS AND DISCUSSION}
Temperature dependent resistivity of Bi single crystal under pressures up to 2.80GPa is shown in Fig.\ref{fig1}. At ambient pressure, Bi adopts the hexagonal structure with semi-metallic property. While, with increasing applied pressure, the resistivity enhances dramatically at low temperatures. An upturn of resistivity in low temperature region appears at 2.2GPa. This pressure induced low-temperature upturn of resistivity was attributed to the semiconductor behavior\cite{SMSC1,SMSC2}. We adopt this picture to interpret our data under low pressures. This semiconducting property becomes more clear at about 2.52GPa. Considering the multiband feature of Bi, the semimetal to semiconductor transition may come from the Fermi level shift of Bi under pressure. Further increasing pressure to 2.63GPa, superconductivity occurs at about 3.9K. Then another superconducting phase with onset temperature of about 7K appears at about 2.80GPa, as shown in the inset of Fig.\ref{fig1}(b). According to the structural phase diagram of Bi under high pressures\cite{Bipressurephase}, the Bi-I phase transforms to Bi-II at about 2.53GPa and Bi-II to Bi-III at about 2.70GPa. Therefore we can probably regard the Bi-II as the superconducting phase with $T_c=3.92$K, and Bi-III as the phase with $T_c=7$K. So the superconductivity induced by pressure is closely associated with structure transitions. As we can see, some other resistivity anomalies are also shown in high temperature region under 2.63GPa and 2.80Ga. These anomalies may be related to the structural transitions of Bi\cite{Bipressurephase}.

\begin{figure}
\includegraphics[width=9cm]{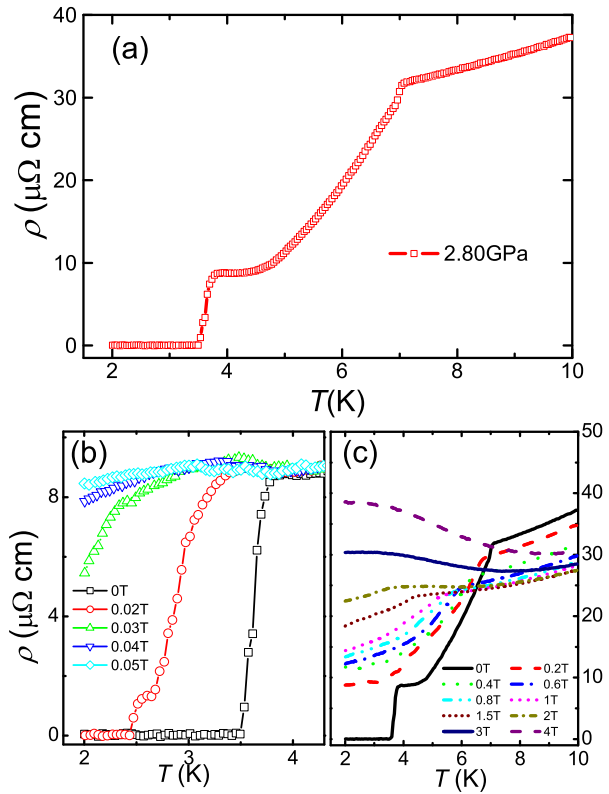}
\caption{(Color online) (a) Enlarged view of the resistive transitions in the temperature range 2K to 10K under 2.8GPa. Two superconducting transitions at 7K and 3.9K are clearly displayed. (b) and (c) The measured resistivity at various magnetic fields of phases Bi-II and Bi-III under 2.80GPa.} \label{fig2}
\end{figure}

In Fig.\ref{fig2}(a), we plot the enlarged view of superconducting transitions at low temperatures at 2.80GPa. Resistivity drops clearly at 7K on the semimetallic background and then a sharp transition to zero resistivity occurs at 3.9K. Both of these two transitions are confirmed as superconducting transition by magnetic measurements in the following sections. Then we measure the resistivity at different magnetic fields to obtain the upper critical fields of each phases. As shown in Fig.\ref{fig2}(b), the Bi-II phase has a quite low upper critical field. A field as low as 0.05T can almost suppress the superconductivity completely above 2K. In sharp contrast, the Bi-III phase has a much higher $\mu_0H_{c2}$, as plotted in Fig.\ref{fig2}(c), although the $T_c$ of Bi-III is only 1.79 times higher than that of Bi-II. With the suppression of superconductivity by external magnetic field, the resistivity shows an upturn behavior again in low temperature region.

\begin{figure}
\includegraphics[width=9cm]{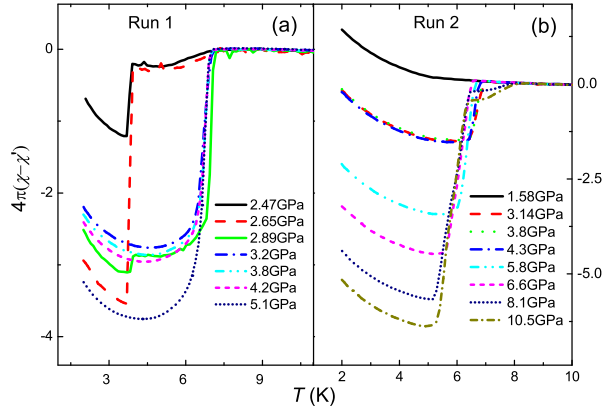}
\caption{(Color online) Temperature dependence of zero-field-cooled (ZFC) magnetic susceptibility for Bi under pressures in two runs of measurements. (a) The measurements of run1 under pressures up to 5.1GPa. Superconductivity firstly emerges at about 4K under 2.47GPa, then the sample totally transforms to the phase with $T_c$=7K under 3.2GPa. (b) The measurements of run2 under pressures up to 10.5GPa. A main superconducting transition occurs at about 7K above 3.14GPa, and another tiny diamagnetic fraction for the phase with $T_c$=8K shows up above 8.1GPa. Due to the large magnetic background of DAC, we first subtract the background of the empty DAC, then we shift all curves to put the final value of magnetic susceptibility at 9K to zero.}
\label{fig3}
\end{figure}

\begin{figure}
\includegraphics[width=9cm]{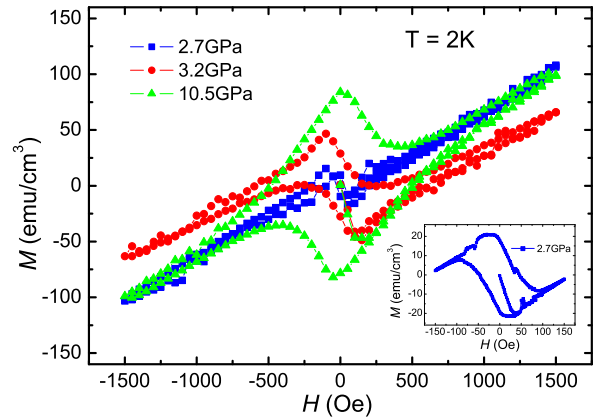}
\caption{(Color online) Magnetization hysteresis loops (MHLs) at 2K under pressures of 2.7GPa (blue square), 3.2GPa (red circle) and 10.5GPa (green trigonal). The inset shows the MHL at 2.7GPa measured in low magnetic field region  considering the small value of upper critical field of phase Bi-II.} \label{fig4}
\end{figure}

In Fig.\ref{fig3}(a), a sharp transition with a large diamagnetic signal can be clearly seen at 3.9K under 2.47GPa, which corresponds to the superconducting transition of phase Bi-II. Meanwhile, a transition with a tiny diamagnetic signal is also visible at about 7K, which may reflect the superconducting transition corresponding to Bi-III. When we increase the pressure to 2.65GPa, the diamagnetic volume of Bi-II becomes much larger than that of 2.47GPa, while the diamagnetic volume of Bi-III phase almost keeps unchanged. Then, at 2.89GPa, Bi-III becomes the dominant superconducting phase and the diamagnetic screening volume of Bi-II becomes smaller, comparing with that of Bi-III. For pressures from 3.2GPa to 5.1GPa, there only exists the Bi-III phase and the $T_c$ slightly decreases with increasing pressure. In Fig.\ref{fig3}(b), we show another run of measurements with pressures up to 10.5GPa. At 1.58GPa, no superconducting transition can be seen, which is consistent with the resistivity measurements. While from 3.14GPa to 6.6GPa, superconducting transition of Bi-III is shown and also the $T_c$ slightly decreases with increasing pressure. According to previous work, Bi-IV phase appears at 5GPa with $T_c$ of about 7K\cite{BipressureSC}. Because this $T_c$ is quite close to that of Bi-III phase, we can not distinguish them from our data. When external pressure increases to 8.1GPa, a superconducting phase with higher $T_c$ (about 8K) appears. Then its diamagnetic volume becomes larger at 10.5GPa and the $T_c$ of this phase gets slightly enhanced. This phase may be recognized as the Bi-VI phase. In order to show the superconductivity is bulk in nature, we present magnetization hysteresis loops (MHLs) in Fig.\ref{fig4} under different pressures. Due to the quite low upper critical field of Bi-II, we further measured the MHLs in the low magnetic field region as shown in the inset of Fig.\ref{fig4}. MHLs clearly demonstrate the bulk superconductivity induced by pressure in bismuth. All of them show typical behaviors of type-II superconductors\cite{shenKFeSe}.

Next, we show the phase diagram of magnetic field versus temperature for Bi-II and Bi-III in Fig.5. The $H_{c2}$ curves are fitted with the simple formula for each phase\cite{ScCoC} as follows

\begin{equation}
H_{c2}(T)=H_{c2}(0)\frac{(1-(T/T_c)^2)^\alpha}{(1+(T/T_c)^2)^\beta}
\end{equation}

\begin{figure}
\includegraphics[width=9cm]{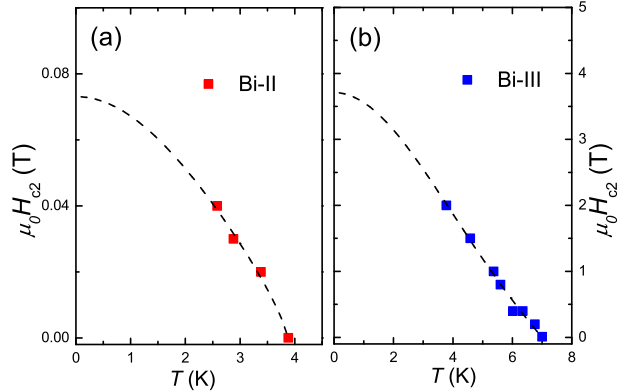}
\caption{(Color online) $H(T)$ phase diagram for Bi-II (a) and Bi-III (b). The dashed black lines are fitting results by using eq.1.} \label{fig5}
\end{figure}

We adopt this expression with two terms of $1-(T/T_c)^2$ and $1+(T/T_c)^2$ as components because they are the basic ingredients for the description of coherence length or upper critical field. According to Ginzburg-Landau theory, $\alpha$=$\beta$=1. We thus fit our data with this general formula. The obtained $\alpha$ and $\beta$ values are 0.79 and 0.47 for Bi-II, 0.99 and 1.04 for Bi-III, respectively. The fitting curves are indicated by the black dashed lines in Fig.\ref{fig5}. As we can see, Bi-II has a very low upper critical field of about 0.073T. This low upper critical field is similar to some other superconducting elements like Tin (3.72K, 0.0308T), Indium (3.40K, 0.0286T) and Tantalum (4.48K, 0.083T)\cite{elementHc}. Comparing to Bi-II phase, Bi-III has a much larger $\mu_0H_{c2}(0)$ of about 3.71T as shown in Fig.\ref{fig5}(b). For type-II superconductors in the dirty limit, $H_{c2}(0)$ could also be given by $H_{c2}(0)=0.691\times\frac{dH_{c2}}{dT}|_{T_c}\times T_c$ \cite{Hc2-1, Hc2-2}. According to this formula, the obtained $\mu_0H_{c2}(0)$ for Bi-II is about 0.103T and 4.56T for Bi-III respectively. These calculated data are both larger than our previous fitting results. It remains to be understood why the Bi-II and Bi-III phases have such different upper critical fields.

\begin{figure}
\includegraphics[width=9.1cm]{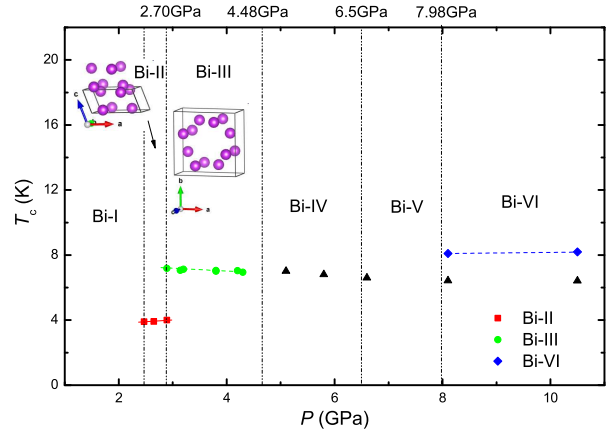}
\caption{(Color online) Phase diagram of superconductivity in Bismuth under pressures. Two schematic pictures indicate the crystal structures of Bi-II and Bi-III. The superconducting transitions around 7K from 5.1GPa to 10.5GPa (black trigonal) can not be distinguished from Bi-III, Bi-IV and Bi-V by our measurements. The Bi-VI phase with $T_c$ of about 8K emerges at about 8.1GPa. The vertical dashed lines are roughly corresponding to the critical pressures of the four sequential structure transitions of Bi under pressures. The colored dashed lines are guidelines for the variation of $T_c$ for each phase. Note the real area for Bi-II occupies only a small region from 2.53 to 2.7GPa.} \label{fig6}
\end{figure}

In Fig.\ref{fig6}, we plot the phase diagram of superconductivity in Bismuth under pressures. Bi starts to become superconductive with $T_c$ of about 3.9K by compressing it to about 2.5GPa, accompanying with a structure transformation from Bi-I to Bi-II. Then, with increasing pressure slightly, Bi-II transforms to Bi-III under 2.8GPa at room temperature. Bi-III also shows superconductivity at about 7K, and $T_c$ displays a small negative slope versus pressure. In the pressure range from 2.47GPa to 2.89GPa, Bi-II and Bi-III coexist, as shown by our magnetization measurements. According to previous reports\cite{BipressureSC}, Bi-IV and Bi-V phases may exist under the pressures from 4.48GPa to 6.50GPa and from 6.50GPa to 7.98GP, respectively. And Bi-IV and Bi-V are both superconductors around 7K although their structures are different. By increasing pressure up to 8.1GPa, Bi-VI phase emerges with superconducting transition temperature at about 8.1K. We must mention that, these structural transitions were determined from the compressibility measurements at room temperature\cite{BiHPphasedetermin}, we cannot rule out the possible mixture or coexistence of meta-stable phases at low temperatures. However, our work can explicitly tell three superconducting phases with distinct transition temperatures. This will stimulate further studies on the interesting multiband system bismuth.

\section{SUMMARY}
In summary, we have systematically investigated superconductivity in bismuth by applying high pressures. Resistivity under pressures up to 2.80GPa reveals that the Bi-I phase has a semimetal to semiconductor transition under pressure and also confirms the superconductivity of Bi-II ($T_c$=3.9K) and Bi-III ($T_c$=7K). Resistivity under different magnetic fields were measured for Bi-II and Bi-III. By fitting the $H_{c2}$ data with a empirical model, the $\mu_0H_{c2}(0)$ for Bi-II and Bi-III are about 0.073T and 3.71T, respectively, which are slightly smaller than those determined by the WHH theory. Magnetic susceptibility measurements give consistent transition temperatures as the resistivity measurements for Bi-II and Bi-III. Beside the superconductivity from Bi-II and Bi-III under low pressures, the Bi-VI phase also emerges with $T_c$ of about 8K at 8.1GPa. Finally, we present the phase diagram of $T_c$ versus pressure of Bi, and relate it with the sequential structure transitions.

\section{ACKNOWLEDGMENTS}
We thank Minghu Fang for stimulating discussions. This work was supported by the Ministry of Science and Technology of China (Grant No. 2016YFA0300404, 2016YFA0401700, 2012CB821403), and the National Natural Science Foundation China (NSFC) with the projects: A0402/11534005, A0402/11190023, 51302133.


\begin{thebibliography}{00}

\bibitem{Biband}S. Golin, Phys. Rev \textbf{166}, 643 (1968).
\bibitem{Bipocket}H. R. Verd\'un and H. D. Drew, Phys. Rev. B \textbf{14}, 1370 (1976).
\bibitem{Bielc-hole}J. P. Michenaud and J. P. Issi, J. Phys. C \textbf{5}, 21 (1972).
\bibitem{BiMR}F. Y. Yang, K. Liu, K. Hong, D. H. Reich, P. C. Searson, and C. L. Chien, Science \textbf{284}, 1335 (1999).
\bibitem{BiSOC}T. Hirahara, T. Nagao, I. Matsuda, G. Bihlmayer, E. V. Chulkov, Yu. M. Koroteev, P. M. Echenique, M. Saito, and S. Hasegawa, Phys. Rev. Lett. \textbf{97}, 146803 (2006).
\bibitem{BiHc2}N. B. Brandt and N. I. Ginzburg, Sov. Phy. JETP \textbf{17}, 326 (1963).
\bibitem{BiAmo}W. Buckel and J. Wittig, Phys. Lett. \textbf{17}, 187 (1965).
\bibitem{BiNano}M. L. Tian, J. G. Wang, J. Kurtz, T. E. Mallouk, and M. H. W Chan, Nano Lett. \textbf{6}, 2773 (2006).
\bibitem{Bisemi}X. Du, S. W. Tsai, D. L. Maslov, and A. F. Hebard, Phys. Rev. Lett. \textbf{94}, 166601 (2005).
\bibitem{BipressureSC}M.A. Il'ina and E. S. Itskevich, J. Exp. Theor. Phys. Lett. \textbf{11}, 218 (1970).
\bibitem{SrMnBi}K. Vinod, A. Bharathi, A. T. Satya, S. Sharma, T. R. Devidas, A. Mani, A. K. Sinha, S. K. Deb, V. Sridharan, and C. S. Sundar, Solid State Commun. \textbf{192}, 60 (2014).
\bibitem{CrAs}W. Wu, J. G. Cheng, K. Matsubayashi, P. P. Kong, F. K. Lin, C. Q. Jin, N. L. Wang, Y. Uwatoko, and J. L. Luo, Nat. Commun. \textbf{5}, 5508 (2014).
\bibitem{MnP}J.-G. Cheng, K. Matsubayashi, W. Wu, J. P. Sun, F. K. Lin, J. L. Luo, and Y. Uwatoko, Phys. Rev. Lett. \textbf{114}, 117001 (2015).
\bibitem{ZrTe5}Y. H. Zhou, J. F. Wu, W. Ning, N. N. Li, Y. P. Du, X. L. Chen, R. R. Zhang, Z. H. Chi, X. F. Wang, X. D. Zhu, P. C. Lu, C. Ji, X. G. Wan, Z. R. Yang, J. Sun, W. G Yang, M. L. Tian, Y. H. Zhang, and H. K. Mao, Proc. Natl. Acad. Sci. U. S. A. \textbf{113}, 2904 (2016).
\bibitem{SMSC1}P. Brown, K. Semeniuk, A. Vasiljkovic, and F. M. Grosche, Phys. Procedia. \textbf{75}, 29 (2015)
\bibitem{SMSC2}N. P. Armitage, R. Tediosi, F. L\'evy, E. Giannini, L. Forro, and D. van der Marel, Phys. Rev. Lett. \textbf{104}, 237401 (2010)
\bibitem{Bipressurephase}S. Yomo, N. M\^ori, and T. Mitsui, J. Phys. Soc. Jpn. \textbf{32}, 667 (1972).
\bibitem{shenKFeSe}B. Shen, B. Zeng, G. F. Chen, J. B. He, D. M. Wang, H. Yang, and H. H. Wen, Europhys. Lett. \textbf{96}, 37010 (2011).
\bibitem{ScCoC}E. Y. Wang, X. Y. Zhu, and H. H. Wen£¬Europhys. Lett. \textbf{115}, 27007 (2016).
\bibitem{elementHc}R. W. Shaw, D. E. Mapother, and D. C. Hopkins, Phys. Rev. \textbf{120}, 88 (1960).
\bibitem{Hc2-1}K. Maki, Phys. Rev. \textbf{148}, 362 (1966).
\bibitem{Hc2-2}E. Helfand and N. R. Werthamer, Phys. Rev. \textbf{147}, 288 (1966).
\bibitem{BiHPphasedetermin}P. W. Bridgman, Proc. Am. Acad. Arts Sci. \textbf{81}, 228 (1952)


\end{thebibliography}
\end{document}